# Hyperbolic Continuous Topological Transition in Real Space

Junke Liao[1], Tao Hou[1], Huanyang Chen[1, *]
1. Department of Physics, Xiamen University, Xiamen 361005, China
*Corresponding author. E-mail: *kenyon@xmu.edu.cn

**Abstract:**

Hyperbolic topological transitions refer to the transformation of is isofrequency contours in hyperbolic materials from one topology (e.g., hyperbolic) to another (e.g., elliptical or a different hyperbolic topology). However, current research remains limited to investigating topological transitions in momentum space, thereby hindering the simultaneous real-space observation of distinct hyperbolic states and their associated topological transitions within a single system. In this work, we investigate real-space hyperbolic continuous topological transitions using gradient-index (GRIN) lenses, exemplified by hyperbolic Luneburg lens. By introducing Wick rotations, we demonstrate how spatially modulated refractive indices, mediated by variations in out-of-plane permittivity, drive continuous transitions between hyperbolic Type I and Type II topologies. Furthermore, using a harmonic oscillator model, we uncover the intrinsic relationship between the parameter $E$ of hyperbolic Luneburg lens and its predominant topological behavior, whether hyperbolic Type I or Type II, and extend this concept to a broader framework of Morse lenses. This work provides a theoretical foundation for designing materials with tunable topological properties, advancing applications in photonics, metamaterials, and beyond.

## 1. INTRODUCTION

Hyperbolic materials (HMs) are characterized by their strong anisotropy, where the diagonal components of the dielectric tensor possess opposite signs, resulting in hyperbolic dispersion in momentum space [1-3]. This unique property enables light or polaritons to propagate within confined regions and specific directions, overcoming conventional optical limitations such as subwavelength focusing and imaging [4,5]. Artificial hyperbolic metamaterials composed of metallic and dielectric elements have been extensively studied in various fields, including negative refraction phenomena [6,7] and nanolithography [8]. In recent years, natural hyperbolic materials like α-$MoO_3$ [9-11] and h-BN [12-16] have garnered significant attention due to their low-loss characteristics and anisotropic optical behaviors in the infrared and terahertz regimes. These materials support applications such as nanoscale waveguides [12,13] and phonon-polariton excitations [9,10, 15,16], providing a novel platform for developing light-matter interactions.

Based on the opening direction of the hyperbolic dispersion relation in momentum space, hyperbolic materials are classified into Type I and Type II, with distinct excitation directions for each type [17-18]. In hyperbolic materials, topologies are of profound significance, as they govern the emergence of unique boundary states [19-20] and enable novel quantum phenomena, including topological insulators, which arise from topological invariants [21-25], and the spin Hall effect resulting from different isofrequency contour topologies [26,27]. These phenomena are critical for advancing quantum imaging, optical computing [28-30] and on-chip integration technologies [31,32]. In recent years, experimental demonstrations of hyperbolic topological transitions (the topological change of isofrequency contours) in momentum space have been achieved, such as tuning the structural stacking angle parameters of van der Waals materials to exhibit polariton-dominated topological states [33], and observing hyperbolic-to-hyperbolic transitions near the exceptional reststrahlen points (ERP) in Y$_2$SiO$_5$ [34]. Although transitions between hyperbolic states have been observed in some systems by varying frequency or altering material structural parameters [33-35], it remains impossible to simultaneously observe different hyperbolic states and their topological transitions within the same system. In particular, studies addressing the coexistence of multiple hyperbolic states in real space and their transitions are still lacking.

In this work, we introduce a gradient-index (GRIN) lens design, exemplified by Luneburg lens, to demonstrate the coexistence of hyperbolic Type I and Type II states and their continuous transition in real space. These transitions are driven by the spatially varying refractive index, specifically the sign change of the out-of-plane permittivity or permeability within the GRIN lenses. By comparing field patterns under different Luneburg parament $E$ values with predictions from harmonic oscillator models, we observe that the system predominantly exhibits characteristics of either Type I or Type II HMs, highlighting a strong dependence on $E$. Furthermore, we derive a unified set of hyperbolic GRIN expressions capable of facilitating continuous topological transitions across regions with differing refractive indices, challenging traditional perspectives and extending the theoretical framework for gradient-index material design.

## 2. RESULTS

The refractive index of a material is determined by the interplay between the in-plane and out-of-plane components of its permittivity and permeability. For a two-dimensional transverse electric (TE) or transverse magnetic (TM) wave, we consider a simplified case where the material profile of the gradient refractive index device is primarily governed by the out-of-plane dielectric parameters:

$$\varepsilon = \mu = \begin{bmatrix} 1 & & \\ & 1 & \\ & & n^2(x,y) \end{bmatrix} \quad (1)$$

The expression for the gradient refractive index device is given by $n^2(x,y)$.

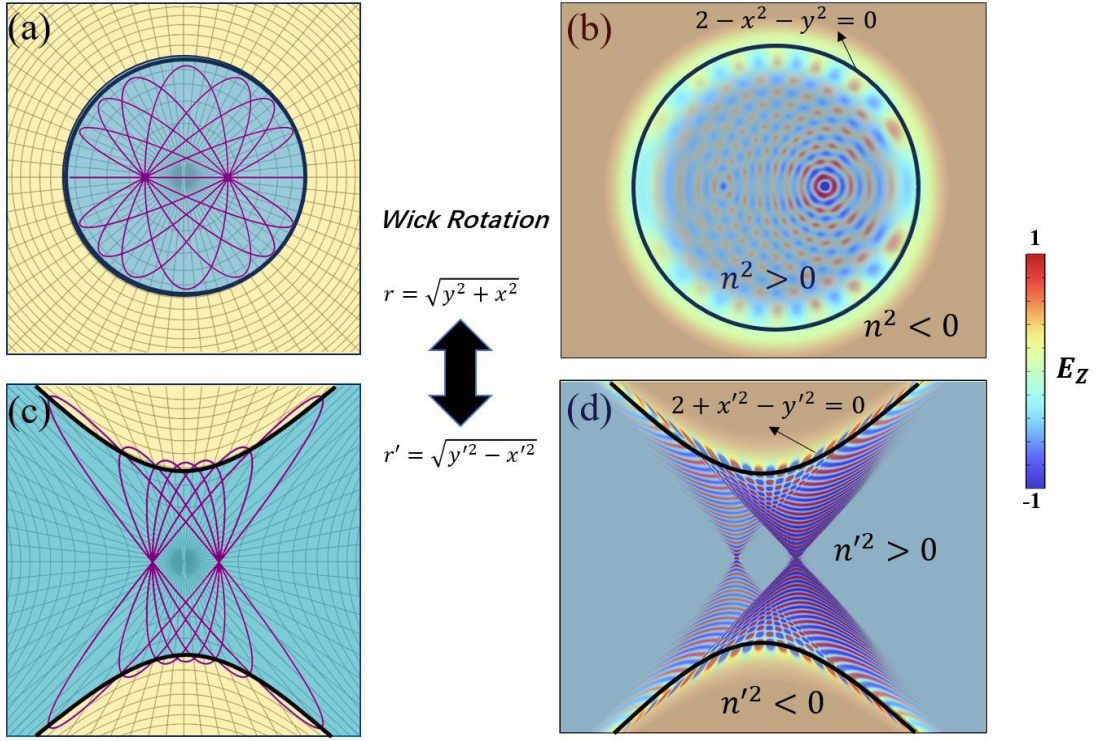

Fig 1. Schematic illustration of the transformation of a Luneburg lens into a hyperbolic Luneburg lens via Wick rotation. (a) and (c) show the trajectories of rays (magenta curves) emitted from the point ($\frac{1}{2}$, 0) within the Luneburg lens and the hyperbolic Luneburg lens, respectively. (b) and (d) present the corresponding electromagnetic TE wave simulations distributions, illustrating the behavior of the systems in (a) and (c). The light ray trajectories are derived using Hamiltonian optics [39,40].

Using Luneburg lens as an example, the refractive index profile is given by $n^2(r) = 2E - r^2$, where $r = \sqrt{x^2 + y^2}$. As shown in Fig. 1(a), the blue region corresponds to $n^2(r)(r \geq \sqrt{2E})$, where electromagnetic waves can propagate, while the yellow region represents the metals or magnetic materials zone outside the lens, where $n^2 < 0 (r^2 \leq \sqrt{2E})$ and electromagnetic waves cannot propagate. The Luneburg lens, as an absolute instrument, is most famously known for its ability to focus electromagnetic waves emitted by a point source to its symmetric position, as shown in Fig. 1(a) and (b). To integrate such a gradient refractive index device into hyperbolic materials, we need to perform a Wick rotation [36]:
$$x' = ix, y' = y \qquad (2)$$
This rotation results in the Minkowski space line element: $ds'^2 = -dx'^2 + dy'^2$, which corresponds to a hyperbolic form of the radial distance, $r' = \sqrt{y'^2 - x'^2}$, as shown in Fig. 1(b). This Wick rotation is essentially a coordinate transformation. According to the transformation optics (TO) [37,38], we can then calculate the refractive index profile of the hyperbolic gradient lens after the Wick rotation:

$$\varepsilon' = \mu' = \begin{bmatrix} 1 & & \\ & -1 & \\ & & n^2(r') \end{bmatrix} \quad (3)$$

where $n^2(r') = 2E + x'^2 - y^2$ for the hyperbolic Luneburg lens. Similarly, when a point source is placed in the hyperbolic Luneburg lens, imaging effects are observed, as shown in Fig. 1(c) and (d). However, an unexpected phenomenon occurs: during the imaging process, electromagnetic waves pass through the boundary where $n'^2(r') = 0$ (indicated by the thick black line) and enter the yellow region characterized by $n'^2(r') < 0$, where electromagnetic waves should take the form of evanescent waves.

The dispersion relation for the two-dimensional Luneburg lens is given by $k_y^2 + k_x^2 = n^2(r)$. For the refractive index profile with $n^2(r) > 0$, after performing a Wick rotation, the dispersion relation becomes $k_y^2 - k_x^2 = n'^2(r')$, corresponding to the Type I hyperbolic material. In contrast, for the refractive index profile with $n^2 < 0$, after the Wick rotation, the dispersion relation takes the form $k_x^2 - k_y^2 = |n'^2(r')|$, corresponding to a Type II hyperbolic material. This indicates that, unlike a conventional Luneburg lens where electromagnetic waves transition from propagating waves to evanescent waves at the boundary, in the hyperbolic Luneburg lens, the wave transition is from one hyperbolic state to another, achieving a continuous topological transition in real space.

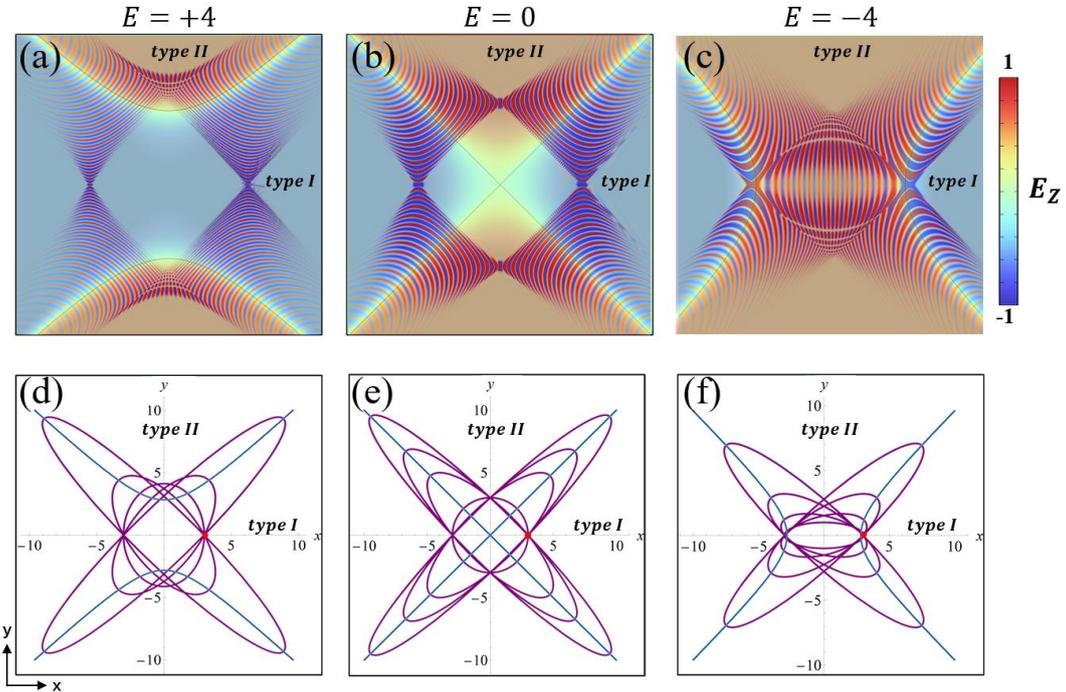

**Fig. 2. Field distributions and ray tracing of hyperbolic Luneburg lenses at different parament E.** Simulation wavelength: λ = 1. (a-c) TE field distributions of the hyperbolic Luneburg lens with a point source located at (3, 0) for energies E = +4, 0, and -4, respectively. The background color represents the refractive index distribution. (d-f) The corresponding ray tracing for the cases shown in panels (a-c). The red dot represents the point source, and the blue line indicates the boundary where $n^2 = 0$.

To provide a more intuitive demonstration of the two distinct types of hyperbolic materials present in the hyperbolic Luneburg lens, and the free propagation of electromagnetic waves within these two types of materials, we analyzed three representative values of the parameter $E$: 0, ±4, as shown in Fig. 2. When $E = 0$, placing a point source on the x-axis of the Luneburg lens, the electromagnetic waves form not only a focal point at the symmetry of the origin but also two additional focal points on the y-axis, exhibiting an C4 symmetry. Therefore, all four points can be considered as sources, as shown in Figs. 2 (b) and (e). For the two points on the x-axis (corresponding to the region with $n'^2(r') > 0$), the wave propagation is confined between the asymptotes, forming the characteristic vertical hourglass shape of Type I hyperbolic materials (with $k_y > k_x$). For the two points on the y-axis (also corresponding to $n'^2(r') < 0$), the propagation is confined such that $k_x > k_y$, exhibiting the horizontal hourglass shape characteristic of Type II hyperbolic materials. Due to the propagation direction restrictions imposed by these two types of hyperbolic materials ($k_y > k_x$ or $k_x > k_y$), a square forbidden region forms in the central area of the hyperbolic Luneburg lens, where electromagnetic waves cannot propagate. For $E = \pm 4$, only two points exist (one focal point and one source point), and the two focal points transform into caustics, as shown in Figs. 2 (d) and (f). However, in the field diagram, we still observe that Type I propagates primarily in the y-direction, and Type II propagates primarily in the x-direction. There is a slight difference in the caustics for $E = \pm 4$. For $E = +4$, the caustics appear outside the square forbidden region, while for $E = -4$, the caustics appear inside it. The transition from focal points to caustic lines driven by the sign of the hyperbolic Luneburg parameter $E$ can be understood through the analogy with a harmonic oscillator model. We draw an analogy between the Wick-rotated Luneburg lens and the Schrödinger equation:

$$-\frac{\hbar}{2m}\left(\frac{\partial^2}{\partial y^2} - \frac{\partial^2}{\partial x^2}\right)\varphi(x,y) + \frac{mw^2}{2}(y^2 - x^2)\varphi(x,y) = E\varphi(x,y) \qquad (4)$$

And then, we employ a separation of variables approach, introducing an integer constant $A$:

$$\begin{cases} \dfrac{\hbar}{2m}\dfrac{\partial^2\varphi(x)}{\partial x^2} - \dfrac{mw^2 x^2}{2}\varphi(x) = -A\,\varphi(x) \\ -\dfrac{\hbar}{2m}\dfrac{\partial^2\varphi(y)}{\partial y} + \dfrac{mw^2 y^2}{2}\varphi(y) = (A+E)\varphi(y) \end{cases} \qquad (5)$$

Although $x$ is transformed to $ix$, the solutions remain in the form of harmonic oscillator wavefunctions.

However, during the solution process, when $\varphi(x)$ is set to a specific eigenstate $\varphi_A(x)$, the corresponding $\varphi(y)$ can only exist in higher-energy states $\varphi_{A+E}(y)$. This results in the electromagnetic waves originating from a point source on the x-axis forming caustics along the y-axis, which appear outside the forbidden zone. When the sign of $E$ is reversed, $\varphi(y)$ is restricted to lower-energy states $\varphi_{A-E}(y)$, causing the caustics to shift into the forbidden zone. Only when $E = 0$ do $\varphi(x)$ and $\varphi(y)$ coexist at the same energy level, producing focused points along the $y$-axis.

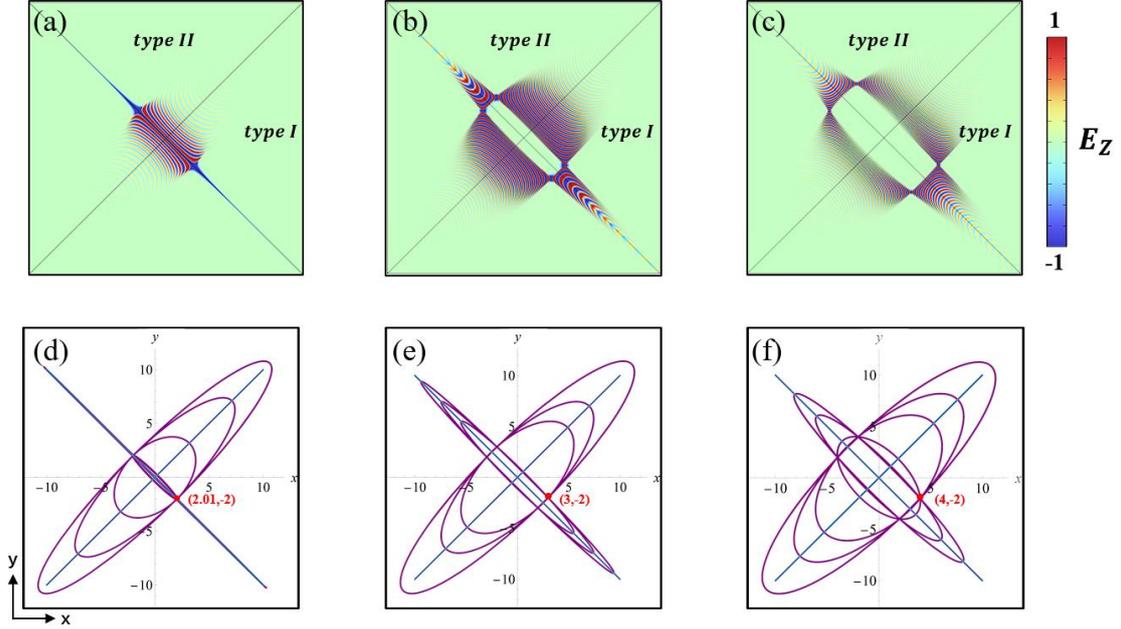

**Fig. 3. Four focal points of the hyperbolic Luneburg lens with E=0 are independent of the point source position.** Simulation wavelength $\lambda = 1$. (a-c) Field distributions for initial conditions corresponding to source positions at (4, 2), (3, 2), and (2.01, 2), respectively. (d-f) The corresponding ray tracing for figures (a-c). The coordinate (2.01, 0) replaces (2, 0) to minimize the short side of the rectangular forbidden zone as it approaches zero.

This indicates that the focusing effect (or caustics) is not influenced by the specific choice of initial coordinates but is solely determined by the parameter $E$. To verify this, we consider the case where $E = 0$ and place the initial condition at a position off the $x$-axis or $y$-axis, such as along $y = \pm x$ or at an arbitrary location, as shown in Fig. 3. For example, when the initial condition is set at $(4, -2)$, the forbidden zone in the central region of the Luneburg lens changes from a square to a rectangle, while the four focal points remain intact. As the distance between the point source and the $y = \pm x$ line is progressively reduced, the shorter side of the rectangular forbidden zone shrinks accordingly. When the short side approaches zero, the four focal points gradually degenerate into two, and the propagation paths of the rays theoretically align with the $y = \pm x$ line. Thus, regardless of the coordinate initial condition, the hyperbolic Luneburg lens with $E = 0$ consistently produces four focal points, a property independent of the point source position.

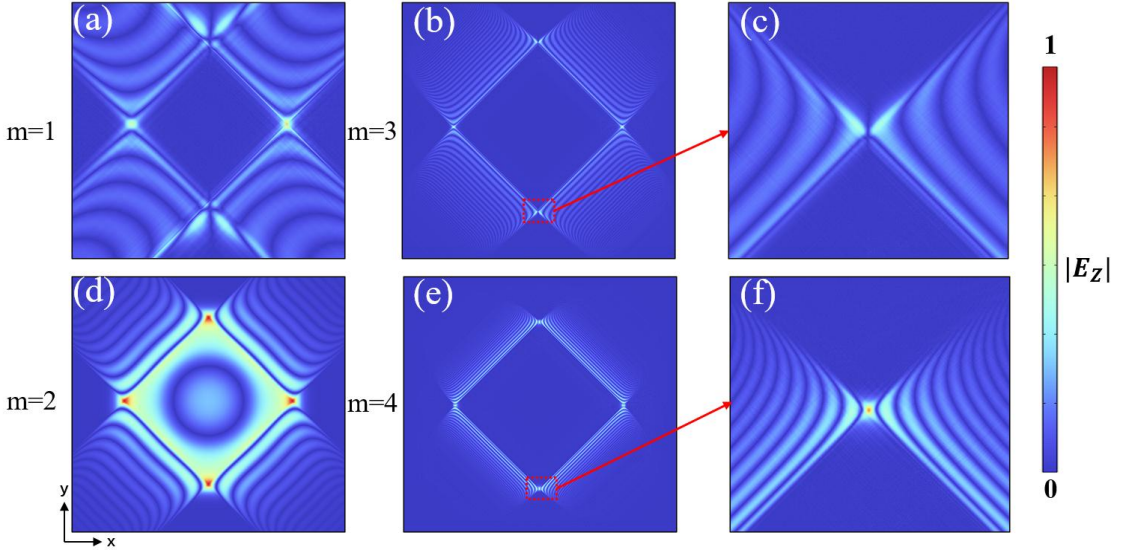

**Fig. 4. Norm of E for hyperbolic Morse lenses of varying orders m.** The simulations are performed at λ=2, with a point source located at (3/2,0). (a-c) Morse lenses of orders 1 and 3, which exhibit a single focal point accompanied by two additional points that form caustics. (d-f) Morse lenses of orders 2 and 4, which produce three focal points without any caustics.

Do all gradient-index lenses exhibit continuous topological transition when Wick rotation is introduced? The answer is evidently no. For instance, prior studies have shown that the Maxwell fish-eye lens cannot support it [41]. This prompts a critical question: which types of gradient-index lenses can sustain hyperbolic continuous topological transition when subjected to Wick rotation? To address this, we extend the concept of the Luneburg lens into a broader framework, referred to as the Morse lens [42,43]. The refractive index profile of Morse lens is described as follows:

$$n^2(r') = \frac{2}{r'^{(a+2)}} - \frac{1}{r'^{(2a+2)}} \qquad (6)$$

where $r' = \sqrt{-x'^2 + y'^2}$. Here, $a = -2$ corresponds to the hyperbolic Luneburg lens. For odd values of $a$, square-root terms emerge, causing undefined regions for $r'^2 < 0$, rendering such configurations unfeasible. For even values of $a \geq 0$, the denominator of the refractive index contains terms like $\frac{1}{r'}$, which diverge as $r' \to 0$, preventing wave propagation. However, when $a = -2m$ ($m = 1,2,3...$), it undergoes a Wick rotation and can achieve hyperbolic continuous topological transitions across boundaries. For $m = 1$, the hyperbolic Morse lens corresponds to the hyperbolic Luneburg lens, as shown in Fig. 4(a). As $m$ increases, as illustrated in Figs. 4(b-d), the electromagnetic field becomes more concentrated, yet the symmetric refocusing property of the hyperbolic Morse lenses remains intact. Thus, each positive integer $m$ corresponds to a unique hyperbolic Morse lens. To generalize this concept, we reformulate Eq. (5) to obtain the expression for this class of hyperbolic lenses:

$$n'^2 = 2r'^{2(m-1)} - r'^{2(2m-1)} \qquad (7)$$

$m = 1,2,3 \ldots$. Here, $r'^{2(2m-1)}$ can take positive or negative values, but $r^{2(m-1)}$ may become non-negative depending on $m$. We observe that:

**1.** When $m$ is **odd**, the non-negativity of $r'^{2(m-1)}$ disrupts the symmetry between positive and negative values of $n'^2$. This asymmetry, as shown in Figs. 4(a-c), results in a single focal point along the $x$-axis, while caustics appear at the $y$-axis.

**2.** When $m$ is **even**, both $r^{2(m-1)}$ and $r^{2(2m-1)}$ can take positive or negative values, preserving the symmetry of $n'^2$. This symmetry, as shown in Figs. 4(d-f), enables the hyperbolic Morse lenses to produce three focal points and eliminating caustics.

This behavior aligns with earlier observations for $E = 0$, where $\varphi(x)$ and $\varphi(y)$ share the same energy levels, maintaining the positive-negative symmetry of $n'^2$. Despite the differences in behavior for odd and even $m$, both cases achieve hyperbolic continuous topological transitions in real space.

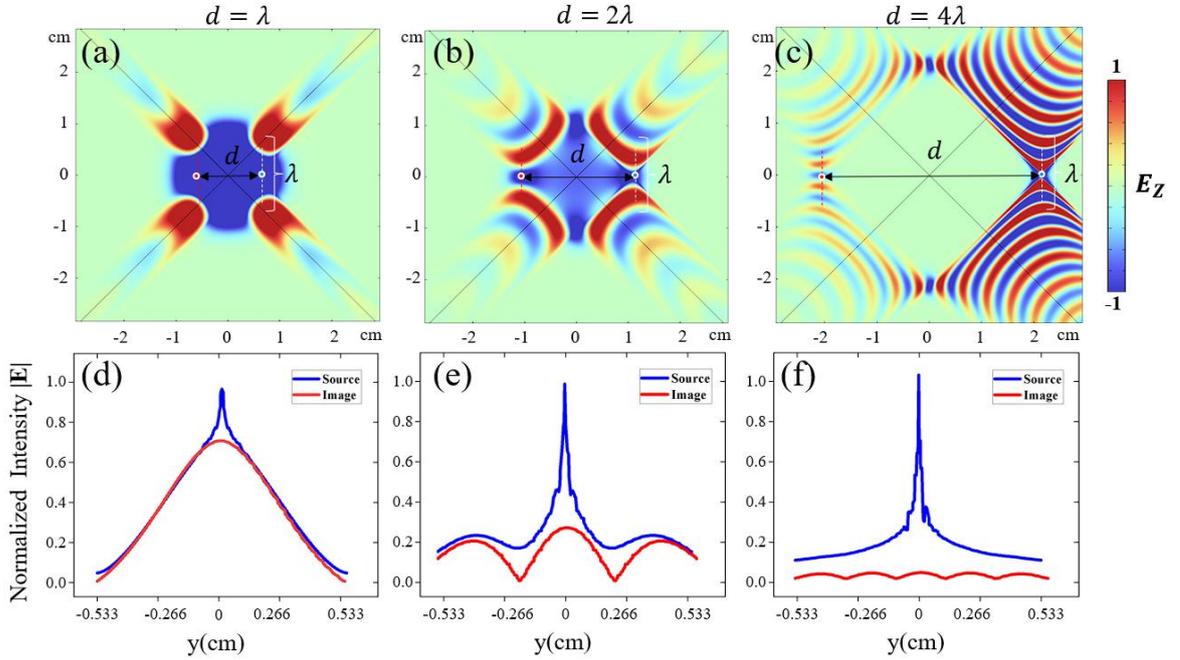

**Fig. 5. Field distribution and corresponding line profiles for hyperbolic Luneburg lenses considering practical losses.** The simulations are performed at $\lambda = 1.063$ cm (corresponding to a frequency of $w = 937$ cm$^{-1}$). The blue dots denote the source points, and the red dots denote the image points. (a–c) Field distributions for source-to-image distances $d=\lambda$, $2\lambda$, $4\lambda$. (d–f) Corresponding normalized intensity profiles |E| along the y-direction (−λ/2 to λ/2) for (a–c).

Although the theoretical framework presented here is largely still in development, recent experimental studies suggest that the proposed transitions could be experimentally achievable. Specifically, α-MoO₃, the recently discovered two-dimensional hyperbolic material, has demonstrated the ability to realize Wick-rotation-induced hyperbolic dielectric profiles. Moreover, Deng et al. [44] have shown how gradient air gap in such two-dimensional hyperbolic material can create gradient index with tunable wavelength ranges. Recent experiments also demonstrate hyperbolic-to-hyperbolic topological transitions in α-MoO₃ by changing its substrate (from SiO₂ to

4H-SiC), further supporting the potential for realizing topological transitions in this work [45]. Therefore, on the α-MoO₃ platform, the combination of a gradient air gap and substrate modification holds great promise for achieving topological transitions in hyperbolic Luneburg lenses. To demonstrate the experimental feasibility, we performed an equivalent simulation by selecting the in-plane permittivity of α-MoO₃ ($\varepsilon_x = 1.377 + 0.025i, \varepsilon_y = -1.374 + 0.098i$) at the frequency of 937 cm⁻¹ [44]. The out-of-plane component was modeled using a gradient air gap to realize the $E = 0$ condition required for the hyperbolic Luneburg lens proposed in this work. Since the wavevector $k$ lies in-plane, the out-of-plane loss can be neglected, and only the real part of the out-of-plane parameter was considered. The simulation results are shown in Fig. 5. We observe that as the distance $d$ between the source and image points increases, the imaging quality gradually deteriorates. This degradation originates from the high in-plane loss of α-MoO₃. Such strong loss would significantly limit the achievable source-to-image distance in practical experiments. However, when the source and image points are very close (d=λ), the hyperbolic Luneburg lens is anticipated to deliver high-quality imaging performance in practical experiments, with the image nearly coinciding with the source, as shown in Figs. 5(a) and 5(d).

## 3. CONCLUSION

In summary, we utilized gradient-index (GRIN) lens designs, including hyperbolic Luneburg lenses and Morse lenses, to demonstrate the coexistence of hyperbolic Type I and Type II states and their continuous transition in real space, marking a significant departure from prior studies focused on hyperbolic topological transitions in momentum space. By introducing Wick rotations, we elucidated how spatial variations in refractive index, driven by the sign change of out-of-plane permittivity or permeability within GRIN systems, govern these transitions. Our analysis of field distributions and ray trajectories under varying parameters $E$ of hyperbolic Luneburg lenses revealed a strong dependence of system behavior on $E$, which determines whether the lens exhibits characteristics of Type I or Type II hyperbolic materials. Additionally, we investigated the refractive index symmetry properties of higher-order Morse lenses and found that asymmetry components significantly influence the number of focal points while still enabling similar hyperbolic continuous topological transitions. This theoretical framework extends the applicability of GRIN materials and provides a foundational basis for controlling real-space hyperbolic topological phenomena. Finally, we analyzed the experimental feasibility using the two-dimensional van der Waals material α-MoO₃. Our findings highlight the potential of GRIN lenses in creating and manipulating hyperbolic structures, thereby broadening the scope of hyperbolic metamaterials and related applications. Furthermore, these results offer new theoretical insights into the generation of phonon polaritons and the manipulation of electromagnetic wave behavior in hyperbolic media, which may also help bridge the concepts of hyperbolic topology in GRIN media and hyperbolic discrete lattices.